\documentclass[prl,twocolumn,showpacs,amsmath,amssymb]{revtex4}

\usepackage[dvips]{graphicx}
\usepackage{graphics}
\usepackage{dcolumn}
\usepackage{bm}

\begin{document}

\title{\begin{center}Control of magnetic anisotropy in (Ga,Mn)As by lithography-induced strain relaxation\end{center}}

\author{J. Wenisch}
\affiliation{Physikalisches Institut, Universit\"{a}t W\"{u}rzburg,
Am Hubland, D-97074 W\"{u}rzburg, Germany}

\author{C. Gould}
\affiliation{Physikalisches Institut, Universit\"{a}t W\"{u}rzburg,
Am Hubland, D-97074 W\"{u}rzburg, Germany}

\author{L. Ebel}
\affiliation{Physikalisches Institut, Universit\"{a}t W\"{u}rzburg,
Am Hubland, D-97074 W\"{u}rzburg, Germany}

\author{J. Storz}
\affiliation{Physikalisches Institut, Universit\"{a}t W\"{u}rzburg,
Am Hubland, D-97074 W\"{u}rzburg, Germany}

\author{K. Pappert}
\affiliation{Physikalisches Institut, Universit\"{a}t W\"{u}rzburg,
Am Hubland, D-97074 W\"{u}rzburg, Germany}

\author{M.J. Schmidt}
\affiliation{Physikalisches Institut, Universit\"{a}t W\"{u}rzburg,
Am Hubland, D-97074 W\"{u}rzburg, Germany}

\author{C. Kumpf}
\affiliation{Physikalisches Institut, Universit\"{a}t W\"{u}rzburg,
Am Hubland, D-97074 W\"{u}rzburg, Germany}

\author{G. Schmidt}
\affiliation{Physikalisches Institut, Universit\"{a}t W\"{u}rzburg,
Am Hubland, D-97074 W\"{u}rzburg, Germany}

\author{K. Brunner}
\affiliation{Physikalisches Institut, Universit\"{a}t W\"{u}rzburg,
Am Hubland, D-97074 W\"{u}rzburg, Germany}

\author{L.W. Molenkamp}
\affiliation{Physikalisches Institut, Universit\"{a}t W\"{u}rzburg,
Am Hubland, D-97074 W\"{u}rzburg, Germany}

\date{\today}

\begin{abstract}

We obtain control of magnetic anisotropy in epitaxial (Ga,Mn)As by
anisotropic strain relaxation in patterned structures. The strain in
the structures is characterized using sophisticated X-ray
techniques. The magnetic anisotropy before patterning of the layer,
which shows biaxial easy axes along [100] and [010], is replaced by
a hard axis in the direction of large elastic strain relaxation and
a uniaxial easy axis in the direction where pseudomorphic conditions
are retained. This strong anisotropy can not be explained by shape
anisotropy and is attributed solely to lattice strain relaxation.
Upon increasing the uniaxial strain anisotropy in the (Ga,Mn)As
stripes, we also observe an increase in magnetic anisotropy.

\end{abstract}

\pacs{75.50.Pp, 75.30.Gw}

\maketitle

The (Ga,Mn)As material system has been the focus of many studies
over the last years. As the understanding of its complex transport
and magnetic properties increases, the focus of interest shifts from
basic research towards its application in devices. For this, it is
necessary to understand how different parameters influence the
ferromagnetic material in structures at the device level. In this
letter we present a systematic study of the role of strain
relaxation as the dominating factor contributing to the magnetic
anisotropy in (Ga,Mn)As nanostructures. A (Ga,Mn)As layer grown
epitaxially on a GaAs substrate is subject to compressive strain in
the plane of the sample and typically exhibits biaxial in-plane easy
axes along [100] and [010], at temperatures around 4~K
\cite{Sawicki, Wang}. In earlier studies, control of the magnetic
anisotropy has been achieved by modifying the strain in the layer.
Tensile strain can be imposed on the (Ga,Mn)As by growing it on a
thick, plastically relaxed (In,Ga)As buffer with a larger lattice
parameter, and results in an out-of-plane easy axis \cite{Liu,
Xiang}.

Here, we follow an alternative approach in modifying the lattice
strain of (Ga,Mn)As on GaAs by lithography, which allows us to
locally control the magnetic anisotropy of the material. By
structuring a fully pseudomorphic 70 nm (Ga,Mn)As layer into thin,
elongated stripes, we allow anisotropic, elastic strain relaxation
perpendicular to the long axis of the stripe. To increase the strain
in the structure compared to the case of (Ga,Mn)As on GaAs, a second
sample is processed which includes a highly compressively strained
layer acting as an extra stressor to the overlying (Ga,Mn)As layer.
The uniaxial strain relaxation in the structures is investigated by
grazing incidence X-ray diffraction (GIXRD) and high-resolution
X-ray diffraction (HRXRD). To determine the influence of patterning
on the magnetic anisotropy, a series of magnetometric and
magnetotransport studies are performed. We also present
finite-element simulations of anisotropic strain relaxation and
\emph{\textbf{k}}$\cdot$\emph{\textbf{p}} calculations which confirm
the relationship between the structural and magnetic behavior
observed in our samples.

The samples are grown in a dedicated III-V MBE chamber with effusion
cells for Ga, In, Mn, and a valved As$_{4}$ cell. A 200~nm thick
high-quality GaAs buffer is deposited at $T_{S}=620$~$^\circ$C, on
an epi-ready semi-insulating (001) GaAs substrate, to improve
surface and layer quality. During a growth interruption, the
temperature is reduced to 270~$^\circ$C. At this temperature, a
(Ga,Mn)As layer of 70~nm thickness is deposited at a rate of
0.86~{\AA}/s and a beam equivalent pressure of As$_{4}$/Ga = 25. The
Mn content is determined by HRXRD to be $\sim2.5$\% \cite{Schott}.
For one sample an additional 80~nm thick (In,Ga)As layer with 7.4\%
In was grown at 500~$^\circ$C prior to (Ga,Mn)As deposition.

\begin{figure}[h]
\includegraphics[angle=0,width=7.5cm]{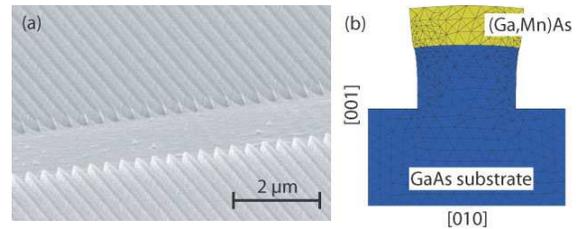}
  \caption{(color online) (a) SEM picture of the sample surface after patterning.
  (b) Simulation of the lattice displacement (100 times exaggerated)
  after strain relaxation in a cross-section of a 200 nm wide stripe.}
  \label{Streifenbild}
\end{figure}

After growth, electron beam lithography and chemically assisted ion
beam etching are used to pattern both samples into the structure
shown in Fig.~\ref{Streifenbild}a. A total area of 4 x 4 mm is
covered by arrays of stripes, each individual stripe measuring
nominally 200~nm x 100~$\mu$m with a separating distance of 200~nm.
The etch depth is about 200~nm, thus well into the GaAs buffer
layer. The stripes are aligned along the [100] crystal direction.

To optimize the stripe dimensions, the relaxation of the stripes has
been modeled by finite element simulations prior to sample
preparation. The simulation makes use of elasticity theory to
describe the strain and displacements in the stripe structure by
minimizing strain energy. A typical resulting cross-section of a
stripe after relaxation is displayed in Fig.~\ref{Streifenbild}b.
Simulations for varied stripe dimensions at a fixed material
composition suggest that a spatially rather homogeneous elastic
relaxation of the (Ga,Mn)As stripe core is reached for stripe widths
below about 500~nm, while inhomogeneous strain relaxation,
concentrated primarily at the edges of the stripe, is observed for
wider structures. The etch depth also plays a critical role in
degree and homogeneity of strain relaxation. As the substrate pillar
height is reduced, the relaxing layer is more restricted at the
lower interface resulting in reduced relaxation and a strain
gradient along the [001] direction.

\begin{figure}
\includegraphics[width=7cm]{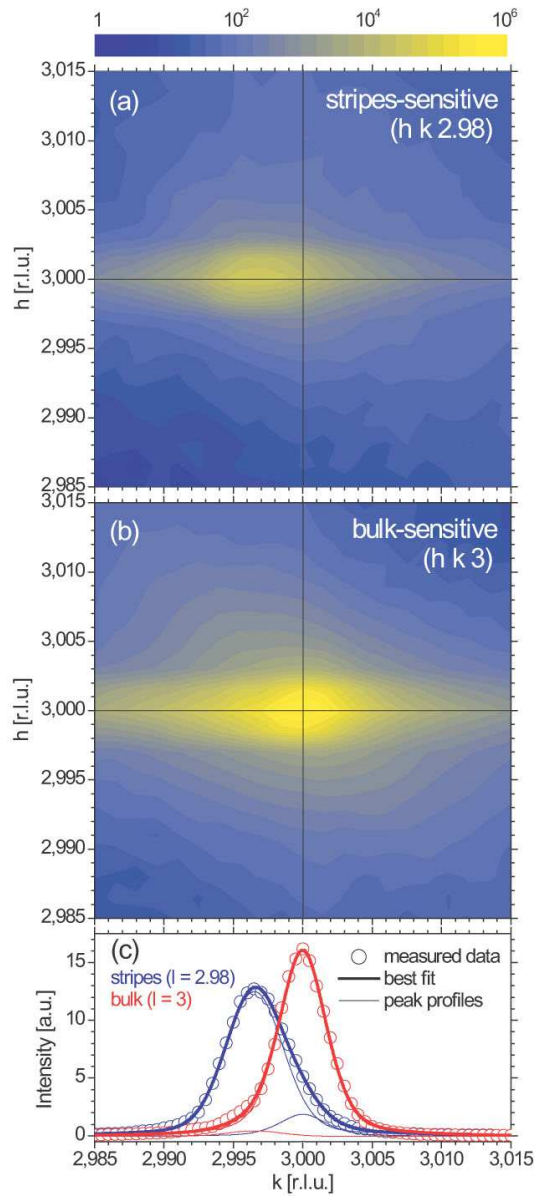}
  \caption{(color online) Reciprocal space
  $h$-$k$-maps in the vicinity of the (333) Bragg reflection at (a) a
  stripes-sensitive and (b) a bulk-sensitive position. In (c) the
  corresponding $k$-line scans (horizontal scans through the maximum
  of (a) and (b)) and best fitting Voigt-profiles are shown, see
  text.}
  \label{fig:map}
\end{figure}

The strain relaxation in the (Ga,Mn)As stripes is determined from
GIXRD reciprocal space mapping in the vicinity of the (333) Bragg
reflection. The GIXRD experiments were performed at beamline BW2 of
the Hamburger Synchrotronstrahlungslabor (HASYLAB). Due to the small
lattice mismatch between GaAs and (Ga,Mn)As of $f=0.15\cdot10^{-3}$,
the Bragg reflections of both materials lie very close to each
other. If the (Ga,Mn)As stripes were purely pseudomorphically
strained, the difference between both reflections would be only
$\Delta l= 0.0033$~r.l.u. (reciprocal lattice units), a value which
can hardly be resolved with our experimental setup. However, since
the thickness of the layer is only about 70~nm, the (Ga,Mn)As (333)
reflection is significantly broadened in $l$-direction. Therefore,
by mapping the reciprocal space close to but not precisely at the
$l$-position of the GaAs bulk-peak one is mainly sensitive to the
(Ga,Mn)As stripes. This measurement, a $h$-$k$-map at $l=2.98$, is
shown in Fig.~\ref{fig:map}a, and a similar map through the GaAs
bulk-peak in Fig.~\ref{fig:map}b. The latter is a bulk-sensitive
measurement and shows a peak located at (333) providing a coordinate
system of the lattice units. In the stripes-sensitive measurement,
one can clearly observe (Fig.~\ref{fig:map}a) a shift of the peak
towards smaller values in $k$. This shift indicates the relaxation
of the (Ga,Mn)As structure in [010] direction, whereas in [100]
direction no relaxation takes place (no peak shift visible in
$h$-direction). The different widths of the peaks in $h$- and
$k$-direction are due to the different lateral dimensions of the
stripes.

In order to quantify the shift, we fit the measured peaks to
Voigt-profiles. Fig.~\ref{fig:map}c shows the central line scan
through the peaks of both reciprocal space maps (open circles in the
figure). Each curve is fit with the sum of two Voigt-peaks (thick
solid lines), one fixed at $h=3$ representing the bulk-contribution,
the other with a variable position. All four individual peaks are
also shown as thin lines. By this procedure we obtain a value for
the strain in [010] direction of
$\epsilon_{[010]}=(a_{meas}-a_{nat})$/$a_{nat}=-2.8\cdot10^{-4}$. We
conclude that the (Ga,Mn)As stripes are pseudomorphically strained
in the [100]- but show a large degree of strain relaxation in the
[010]-direction.

\begin{figure}
\includegraphics[angle=0,width=7.5cm]{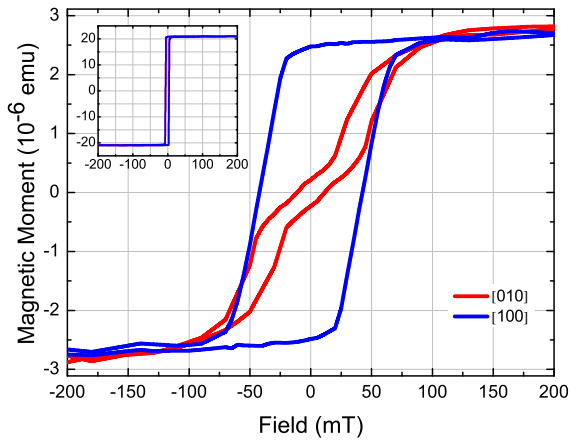}
  \caption{(color online) SQUID magnetization data at 4 K for
  the parent layer (inset) and the patterned layer with stripes aligned
  along [100].}
  \label{SQUID}
\end{figure}

For magnetic characterization, a superconducting quantum
interference device (SQUID) is used to measure magnetization
behavior along selected directions in the parent and patterned
layers. The results are shown in Fig.~\ref{SQUID}.

The parent layer (inset of Fig.~\ref{SQUID}) exhibits the well known
biaxial anisotropy at 4~K with easy axes along the [100] and [010]
crystal directions and hard axes along [1$\bar{1}$0] and [110]. The
two easy axes are both plotted but not distinguishable in the inset.
Fig.~\ref{SQUID} shows the anisotropy after patterning. Along the
[010] direction, the magnetic moment at zero field drops by
$\sim90$\% compared to the [100] direction. This previously easy
axis is now clearly a very hard magnetization axis. The remaining
easy axis along [100] also reflects this change by an increase in
coercive field from 4~mT in the parent layer to 43~mT in the
patterned layer. This effect is caused by the increased energy
necessary to rotate the magnetization through the now hard [010]
axis during magnetization reversal. Thus, the presented results show
a clear modification of the magnetic anisotropy of nanopatterned
(Ga,Mn)As/GaAs stripes.

The observed strong anisotropy can not be explained by shape
anisotropy. As-grown (Ga,Mn)As shows mainly biaxial crystalline
magnetic anisotropy ($K_{cryst}\sim3000$~J/m$^{3}$), which dominates
the magnetic reversal behavior. Calculating the uniaxial term
stemming from the shape anisotropy produced by our geometry
(ferromagnetic prism \cite{Aharoni}) leads to
$K_{shape}\sim280$~J/m$^{3}$, much smaller than the crystalline
anisotropy contribution. Thus we attribute the observed emergence of
the dominant uniaxial anisotropy in our samples to the anisotropic
strain relaxation which, as we show in the
\emph{\textbf{k}}$\cdot$\emph{\textbf{p}} modeling below,
contributes an additional term in the magnetostatics energy
equation, with enough impact to overcome the otherwise dominant
biaxial anisotropy.

\begin{figure}
\includegraphics[angle=0,width=8cm]{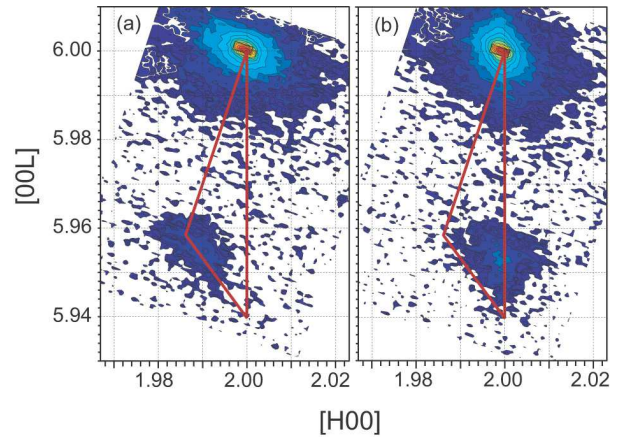}
  \caption{(color online) Reciprocal space maps of the (206) Bragg reflection with incident
   X-rays (a) perpendicular and (b) parallel to the stripe axis. The
   solid red line indicates the triangle of relaxation for this special case of uniaxial relaxation.}
  \label{206RSM}
\end{figure}

It is possible to further increase the strength of the imposed
uniaxial anisotropy by increasing the strain in the parent layer.
For this we use an (In,Ga)As layer with considerably larger lattice
mismatch of $f=0.53\cdot10^{-3}$ to the substrate, deposited prior
to the (Ga,Mn)As layer with $f=0.15\cdot10^{-3}$. This layer, which
is grown to a thickness close to but below the critical value for
plastic strain relaxation, acts as a stressor to the overlying
(Ga,Mn)As layer. In Fig.~\ref{206RSM} we present HRXRD data of the
patterned stripe structure. To distinguish between the lattice
constants along the long axis of the stripes ([100] direction) and
perpendicular to the stripes ([010] direction), two reciprocal space
maps (RSM) of the (206) Bragg reflection were taken with the
incident X-ray beam perpendicular (Fig.~\ref{206RSM}a) and parallel
(Fig.~\ref{206RSM}b) to the stripe axis. In the first case, the
(In,Ga)As layer peak maximum clearly indicates a larger lateral and
smaller vertical lattice constant than expected for a fully
pseudomorphic layer (bottom tip of the relaxation triangle). The
(Ga,Mn)As layer is expected to have a similar lateral lattice
constant and therefore lateral peak coordinate in the RSM to the
underlying (In,Ga)As. Hence, the bulge in the substrate peak at
(1.992, 5.988) and (2.000, 5.986) in Fig.~\ref{206RSM}a and
Fig.~\ref{206RSM}b, respectively, is attributed to the (Ga,Mn)As
layer. Because of the peak location, we conclude that tensile strain
($\epsilon=2.2\cdot10^{-3}$) has been induced by the relaxed
(In,Ga)As layer. With incident X-rays along [100]
(Fig.~\ref{206RSM}b), both layer peaks are located directly below
the substrate peak, but shifted to smaller vertical lattice constant
(due to Poisson effect) compared to the biaxial pseudomorphic case.
The position of the peaks in both maps can be explained
unambiguously by a large uniaxial strain relaxation in the [010]
crystal direction while retaining fully pseudomorphic conditions
along the [100] direction.

This sample shows a very similar behavior in SQUID compared to the
pure (Ga,Mn)As sample, with the well-known biaxial easy-axes in the
parent layer, as well as the emergence of the strong uniaxial
anisotropy after patterning.

\begin{figure}
\includegraphics[angle=0,width=7.5cm]{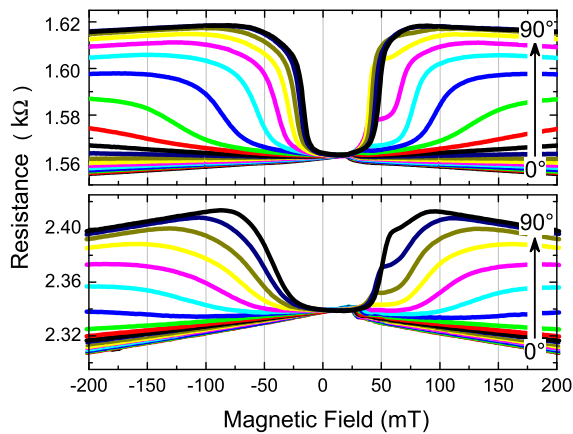}
  \caption{(color online) Magnetoresistance scans of (Ga,Mn)As/GaAs (top) and (Ga,Mn)As/(In,Ga)As/GaAs sample (bottom) at 4 K for
  angles $\phi$ between magnetic field and current direction (along stripe) from 0$^\circ$ to 90$^\circ$. The magnetic field is
  swept from $-0.3$ T to 0.3 T for each scan.}
  \label{transport}
\end{figure}

That the imposed uniaxial anisotropy in the sample containing the
(In,Ga)As layer stressor is even stronger, can be inferred from
magnetoresistance measurements. In (Ga,Mn)As, which is inherently
highly p-doped by Mn, a strong anisotropic magnetoresistance (AMR)
is generally observed. The magnetoresistance is dependent on the
angle $\phi$ between the magnetization $\vec{M}$ and the current
$\vec{J}$ \cite{McGuire}, with a minimum when $\vec{M}$ is parallel
to the current \cite{Baxter}. For the measurement (see
Fig.~\ref{transport}), about 250 parallel stripes are contacted from
both ends. A series of magnetic field sweeps from $-0.3$~T to 0.3~T
is performed, with incremental increase of the angle $\phi$ between
$\vec{M}$ and $\vec{J}$ after each sweep. In both samples, all
(hysteretically symmetric) curves share one low resistance state at
$B=0$~T. When the field sweep is performed in the [100] direction
($\phi=0^\circ$), the resistance remains at the low state. With
increasing angle between $\vec{M}$ and $\vec{J}$, a high resistance
state develops and reaches a maximum at $\phi=90^\circ$. At this
angle, the magnetization starts out parallel to the stripes (low
resistance) and is forced to the unfavorable direction perpendicular
to the stripes (high resistance) at high fields. As with the SQUID
results, these findings fit perfectly into the picture of a strain
induced uniaxial magnetic anisotropy. We also observe a larger
opening of the curves from the additionally strained stripes (lower
part of Fig.~\ref{transport}). This is a direct evidence of an
increased anisotropy field, as more energy is necessary to rotate
the magnetization into the unfavorable hard [010] direction.
According to \cite{nanobars}, we determine the magnitude of the
anisotropy field from these openings to be $\sim45$~mT for the pure
(Ga,Mn)As stripes and $\sim80$~mT for the additionally strained
stripes. As the strain in the magnetic layer of the second sample is
demonstrably significantly different with nominally identical
physical dimensions, this effect can clearly be attributed to the
lattice strain, considering that shape anisotropy is identical for
both structures.

\begin{figure}
\includegraphics[angle=0,width=7.5cm]{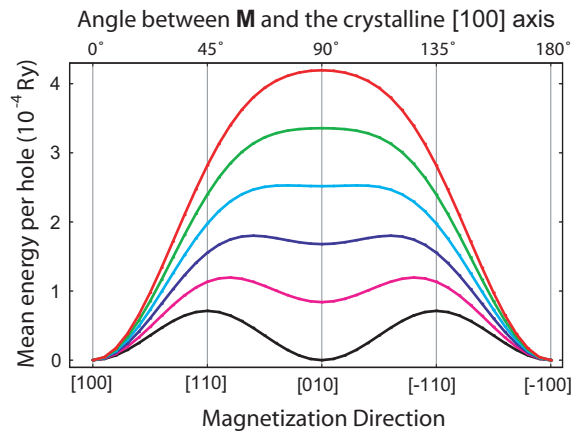}
  \caption{(color online) Mean energy per valence band hole of a (Ga,Mn)As layer with various levels
  of lattice strain in [010] direction ranging, in equal steps, from the biaxial pseudomorphic
  case ($\epsilon=-1.5\cdot10^{-3}$, bottom, black) to the fully relaxed case ($\epsilon=0$,
  top, red). The strain in [100] remains fixed at $\epsilon=-1.5\cdot10^{-3}$.}
  \label{calculations}
\end{figure}

Finally, we present the results of
\emph{\textbf{k}}$\cdot$\emph{\textbf{p}} calculations \cite{Dietl}
of the magnetization-direction dependence of the mean energy per
hole. The results support our interpretation that strain is the key
element in determining the magnetic anisotropy in (Ga,Mn)As
nanostructures. Fig.~\ref{calculations} shows the evolution of the
equivalent energy minima along [100] and [010] for fully biaxially
strained (Ga,Mn)As (black curve) with increasing degree of strain
relaxation in the [010] direction. Assuming homogeneous strain
distribution in the stripes and a moderate carrier density
$4\cdot10^{20}$~cm$^{-3}$, we can deduce from the calculations, that
the uniaxial contribution from strain dominates over the biaxial
term to such a degree, that above $\epsilon=-0.6\cdot10^{-3}$ only a
single stable point, along [100], remains. This is characteristic of
perfect uniaxial behavior. As we demonstrated above in connection
with the X-ray data, our patterned samples exhibit strain levels
that exceed this number. The difference by a factor of 2 in the
anisotropy fields for the two differently strained samples
(Fig.~\ref{transport}) also agrees well with the model prediction
(we find a factor of 3). We conclude that our experiments and
modeling unambiguously demonstrate that lithographic pattering, when
necessary combined with the (In,Ga)As stressor technique, is a
reliable and versatile tool for controlling the magnetic
anisotropies in (Ga,Mn)As. We anticipate that this local technique
should prove very useful for device applications.

We thank T. Borzenko, A. Stahl, I. Gierz, and E. Umbach as well as
the staff at HASYLAB for their help in this work. We acknowledge
financial support from the DFG (BR 1960/2-2 and SFB 410) and the EU
(NANOSPIN FP6-IST-015728 and the IHP programme ``Access to Research
Infrastructures``).

\end{document}